\documentclass[conference]{IEEEtran}
\IEEEoverridecommandlockouts
\usepackage{cite}
\usepackage{comment}
\usepackage{amsmath,amssymb,amsfonts}
\usepackage{algorithmic}
\usepackage{graphicx}
\usepackage{textcomp}
\usepackage{xcolor}
\usepackage[pscoord]{eso-pic}

\usepackage{multirow}
\usepackage{siunitx}

\def\BibTeX{{\rm B\kern-.05em{\sc i\kern-.025em b}\kern-.08em
    T\kern-.1667em\lower.7ex\hbox{E}\kern-.125emX}}
\def\BibTeX{{\rm B\kern-.05em{\sc i\kern-.025em b}\kern-.08em
    T\kern-.1667em\lower.7ex\hbox{E}\kern-.125emX}}

\usepackage{fancyhdr}
\fancypagestyle{mahmood}{%
  \fancyhf{} 
  
  \fancyhead[C]{\footnotesize \textcopyright 2020 IEEE. Personal use of this material is permitted.  Permission from IEEE must be obtained for all other uses, in any current or future media, including reprinting/republishing this material for advertising or promotional purposes, creating new collective works, for resale or redistribution to servers or lists, or reuse of any copyrighted component of this work in other works.}
}%
\makeatletter
\let\ps@IEEEtitlepagestyle\ps@mahmood
\makeatother

\begin{document}
\title{InfiniWolf: Energy Efficient Smart Bracelet for Edge Computing with Dual Source Energy Harvesting\\
}
\vspace{-1.55cm}

\author{\IEEEauthorblockN{Michele Magno\IEEEauthorrefmark{2}\IEEEauthorrefmark{3}, Xiaying Wang\IEEEauthorrefmark{2}, Manuel Eggimann\IEEEauthorrefmark{2}, Lukas Cavigelli\IEEEauthorrefmark{2}, Luca Benini\IEEEauthorrefmark{2}\IEEEauthorrefmark{3}}
\IEEEauthorblockA{\IEEEauthorrefmark{2}Dept. of Information Technology and Electrical Engineering, ETH Z\"{u}rich, Switzerland} \IEEEauthorblockA{\IEEEauthorrefmark{3}Dept. of Electrical, Electronic and Information Engineering, University of Bologna, Italy}
\vspace{-0.75cm}
}
\maketitle

\vspace{-1.25cm}
\begin{abstract}
This work presents \emph{InfiniWolf}, a novel multi-sensor smartwatch that can achieve self-sustainability exploiting thermal and solar energy harvesting, performing computationally high demanding tasks. The smartwatch embeds both a System-on-Chip (SoC) with an ARM Cortex-M processor and Bluetooth Low Energy (BLE) and \emph{Mr. Wolf}, an open-hardware RISC-V based parallel ultra-low-power processor that boosts the processing capabilities on board by more than one order of magnitude, while also increasing energy efficiency. We demonstrate its functionality based on a sample application scenario performing stress detection with multi-layer artificial neural networks on a wearable multi-sensor bracelet. Experimental results show the benefits in terms of energy efficiency and latency of Mr. Wolf over an ARM Cortex-M4F micro-controllers and the possibility, under specific assumptions, to be self-sustainable using thermal and solar energy harvesting while performing up to 24 stress classifications per minute in indoor conditions. 
\end{abstract}


\section{Introduction}
Technology advancements in low-power integrated circuits (IC), sensors and wireless transmission have enabled the design and implementation of light-weighted and unobtrusive devices \cite{dang2019novel}. A successful class of such devices is wearable devices, where electronics are worn on the human body. Wearable devices a commercial reality are today with many products for fitness and entertainment applications already being available on the market \cite{jo2019there}. However, one of the most critical issues of those devices is the limited energy supply due to the small size of the battery. 

Machine learning (ML) is playing a crucial role in many fields, including wearable devices used for biomedical applications \cite{pascual2019self, forooghifar2018self}. ML is used for data analysis of the sensor data, and they are the core perception units in our integrated intelligence era \cite{metz2017google}. The new trend is achieving ``smart wearables'' that can extract useful information from several novel sensors attached to the body with \emph{in situ} processing. Thus, embedded ML algorithms are expected to enable machine intelligence in resource-constrained wearable devices~\cite{magnoiwasi2017}. To achieve this goal, researchers working on embedded ML are designing specialized hardware architectures to deal with the demand for large computational and storage capability \cite{annese2017wearable, bernabe2017parallel, rodriguez2018fpga}. 
Due to the computational effort and large datasets, embracing typical ML methods for machine intelligence is still a challenging task in battery-powered wearable devices with limited hardware capabilities~\cite{forooghifar2018self, pascual2019self}. Typically, processors for wearable devices are microcontrollers, often from the ARM Cortex-M family, achieving few million operations per second (MOPS) with a power consumption of a few mW, compatible with the battery of the wearable device. On the other hand, the computational resources of microcontrollers tend to be insufficient to run ML algorithms on the sensor data. Recently, researchers have proposed novel processing units to overcome those limits, enabling on-board data processing with state-of-the-art ML algorithms \cite{conti2016accelerated} . The most promising one exploits parallelism as much as possible, achieving extremely high energy efficiency with near-threshold transistor operation~\cite{pullini2018}. 
However, low-power and energy-efficient processors are not enough to achieve the ultimate goal of wear-and-forget. This ambitious goal is achievable only if the device harvests enough energy from its environment. Harvesting energy from environmental sources has been explored in a wide range of application scenarios. However, in a small-sized wearable context, it is still a challenging task due to the restrictive form-factor constraints~\cite{magno2017wearable}. Steady progress is being made in this area: traditional ``timekeeping only'' watches are fully harvesting powered, and several mature products are available on the market. On the other hand, smartwatches need much more power for their much more sophisticated sensing capabilities and functions. Moreover, harvesting energy to supply always-on wearable devices represents an exciting challenge, which needs particular design attention. Environmental energy sources vary significantly over time, and power management needs to implement methods that can opportunistically take advantage of periods of overabundant energy and survive intervals when the system is starving for energy.

This paper proposes \emph{InfiniWolf}, a smartwatch that exploits the most advanced hardware and software low-power design in combination with power management and energy harvesting technologies to achieve a perpetual and endlessly wearable smart sensor-rich device. In particular, we integrate advanced RISC-V based parallel ultra-low power (PULP) open-hardware processor coupled with aggressive power management and highly efficient environmental energy conversion into a single wearable system. The main goal is to shape an entirely new class of smartwatches that outpaces the state-of-the-art in energy autonomy ready to run sophisticated artificial intelligence algorithms. We show the superior energy efficiency achieved by the developed device and the RISC-V based processor over an ARM Cortex-M4F micro-controller. To evaluate the performance of the implemented smartwatch, we present a challenging application use case of stress detection using a multi-layer perceptron (MLP), a very well-known class of artificial neural networks (ANNs). 

\section{System Architecture}
\begin{figure}[!t]
	\centering
	\includegraphics[width=\columnwidth]{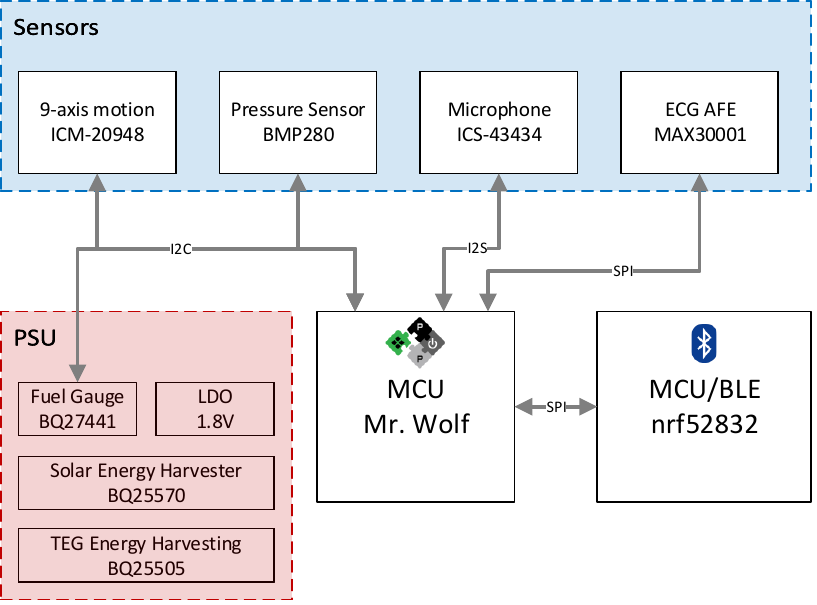}
	\vspace{-0.55cm}
	\caption{Block diagram of InfiniWolf and the smart power unit that is able to harvest energy from dual sources.}
	\vspace{-0.35cm}
	\label{fig:block_diagram}
\end{figure}

Fig.~\ref{fig:block_diagram} shows the block diagram and the architecture of the designed smartwatch. The smartwatch features two processors, a Nordic nRF52832 Bluetooth low energy SoC with an ARM Cortex-M4 processor and a programmable Parallel Ultra-Low-Power processor called Mr. Wolf. The Nordic SoC handles communication with a remote host and auxiliary support for necessary data processing if needed. It provides Bluetooth Low Energy (BLE) 5 communication capabilities, it performs power management various modes of operation (sleep, raw data streaming, data acquisition, and processing) and keeps track of the battery charging status. Mr. Wolf offers high versatility and high compute efficiency at an ultra-low-power budget~\cite{pullini2018}. It features two power domains: SoC and Cluster. The former contains a tiny (12\,kGate) RISC-V processor~\cite{schiavone2017} called \emph{Ibex} and the latter is equipped with 8 RISC-V cores with custom DSP extensions called RI5CY~\cite{schiavone2017}. The dual-processor architecture allows local end-to-end processing (i.e., on-board classification using ML) with lower power and higher energy efficiency than streaming the data out for remote analysis. Moreover, this architecture allows lower latency and more robustness with respect to wireless connectivity.

\begin{figure}[!t]
	\centering
	\includegraphics[width=\columnwidth]{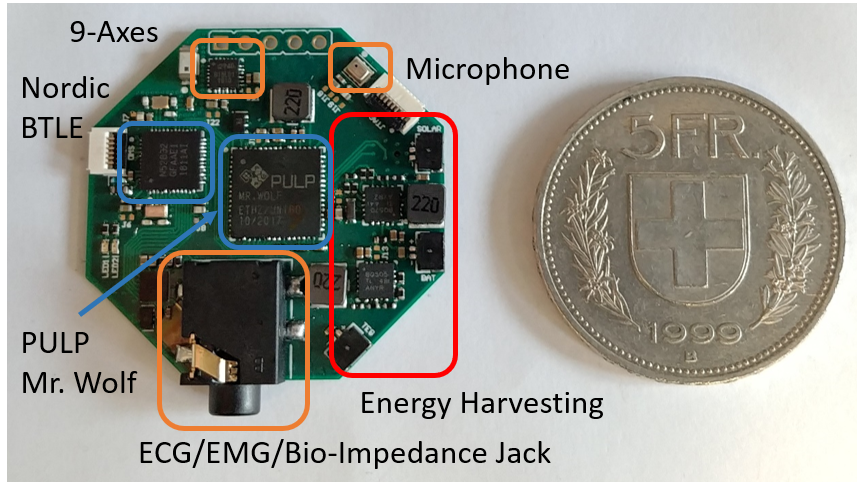}
	\vspace{-0.85cm}
	\caption{InfiniWolf prototype used to carry out experimental measurements.}
	\vspace{-0.55cm}
	\label{fig:board}
\end{figure}

A dual-source energy harvester for solar and thermoelectric generator (TEG) modules has been embedded. The main goal is to achieve perpetual work when the energy transducers are deployed on a wrist band harvesting energy from light and body heat. The choice of two energy harvesting sources is motivated by increased flexibility and robustness of the energy intake, while form-factor is not compromised as the two harvesters exploit different sides of the watch (top for solar and bottom side for thermal). The smart power supply unit (PSU) allows the system to operate with low losses while harvesting energy, monitoring sensors and managing the power according to the policies implemented. The smartwatch includes a 9-axis motion sensor, a pressure sensor, a microphone, and an ultra-low-power ECG/EMG and bioimpedance analog front-end (Maxim MAX30001) to acquire biomedical signals as well as a low power galvanic skin response (GSR) front-end. The wearable device can be worn on the user's wrist and periodically and opportunistically acquires information from the sensors according to the available energy. The device is holistically designed to be ultra-low power and without any need to replace or manually recharge the on-board 120\,mAh LiPo battery. Thus, the design allows extreme power management to keep the quiescent and operative energy consumption low. Figure~\ref{fig:board} shows the smartwatch prototype used to measure data on power consumption and energy efficiency.

\section{Multi-Layer Perception and Stress Detection}
 
In the last few decades, ANNs have been widely used to provide machines with the human ability to make decisions, detecting patterns, predicting actions. An MLP is a class of feed-forward ANNs which comprises at least three layers of neurons: an input layer, a hidden layer, and an output layer. Despite the developments of other types of ANNs such as convolutional and recurrent neural networks, MLPs are still very commonly used due to their simplicity and good performance in many tasks such as emotion classification~\cite{magnoiwasi2017} and disease detection~\cite{wan2018}.

The main goal of implementing a real-world application is to provide an insightful performance comparison between the two processors present in the developed smartwatch, namely an ARM Cortex-M4 and the PULP-based Mr. Wolf. We used the dataset in~\cite{dataset} that includes labeled data of stressed and unstressed people. Previous work~\cite{dataset, alic, chui, bakker} evaluated the most relevant features for the task. However, when it comes to battery-operated wearable devices one has to take the resource constraints of the microcontroller into consideration, especially its limited memory. The work in~\cite{fanncortexm} extracts features and designs an MLP taking into account these memory constraints. Based on~\cite{fanncortexm}, we obtained 5 relevant features from ECG and GSR data and trained an MLP to perform stress detection. The dataset is split into subsets with equal stress levels (i.e., transitions between different stress levels are omitted) and extract features by using overlapping windows. From the ECG data, we derived the root mean square (RMSSD) and the standard deviation (SDSD) of the differences of successive neighboring RR intervals, and the number of adjacent RR interval pairs that differ by more than 50\,ms (NN50). For the GSR signals, we detect the rising edges and calculate the height (GSRH) and the length (GSRL) of the slope following~\cite{bakker}.

\begin{figure}[t]
	\centering
	\includegraphics[trim={0.3cm 1.0cm 8cm 2.5cm},clip=true, width=0.8\columnwidth]{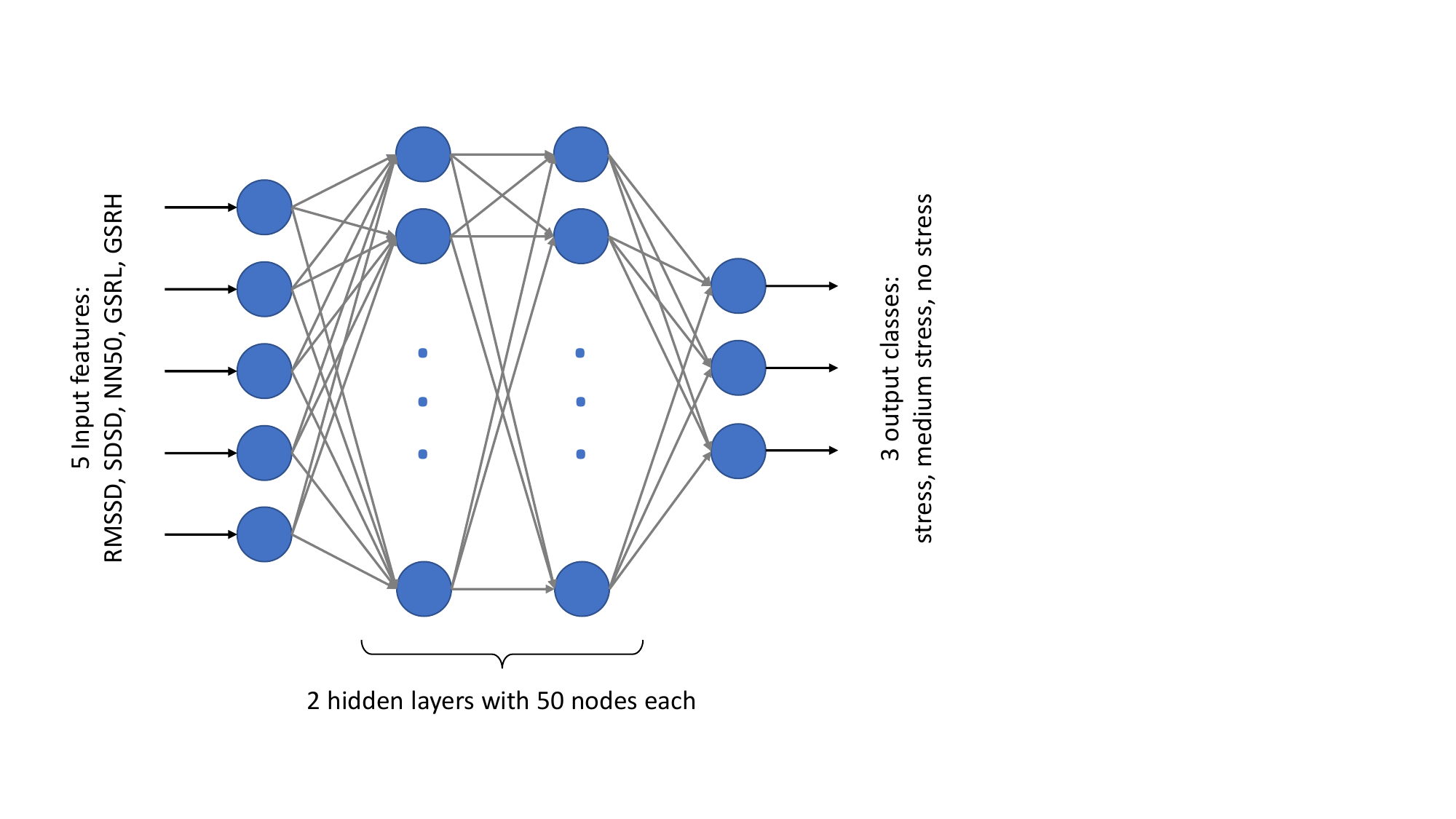}
	\vspace{-0.55cm}
	\caption{Network architecture designed for classifying 3 levels of stress.}
	\vspace{-0.45cm}
	\label{fig:mlp}
\end{figure}

We trained an MLP using the FANN library~\cite{fannlib}, feeding these 5 features as input. The network consists of 2 hidden layers with 50 units each and 3 output nodes in the output layer as shown in Fig.~\ref{fig:mlp}. Here we name it Network A. The used activation function is the tanh function. The network has in total 108 neurons and 3003 weights, yielding an estimated memory footprint of 14\,kB. 
Each neuron requires 4 integers, i.e. 16 bytes, to store the information such as the activation function used, the neuron indexes, etc. Each weight uses 4 bytes of storage and two additional integers are stored per layer, i.e. 8 additional bytes, to save the number of inputs and outputs for each layer. To further demonstrate the energy efficiency of our proposed platform, we designed a bigger network, named Network B, which still fits into the memory and measured the performance. This bigger network has 100 input and 8 output nodes and contains 24 hidden layers with an increasing number of nodes, i.e. the first two hidden layers have 8 neurons each, the subsequent couple of hidden layers have 8 more neurons each, and so on, with a total of 1356 neurons, 81032 weights and an estimated memory size of 353\,kB. To get consistent measurements, we repeat all experiments 1000 times with a warm cache and then average the results.

\section{Preliminary Experimental Results}
This section shows the experimental results on the execution time and energy of InfiniWolf to perform stress detection. We measured the energy harvested by the two transducers to evaluate the possibility to achieve self-sustainability. Moreover, we present a comparison between two processors: Mr. Wolf and ARM Cortex-M4F on the Nordic nRF52832, running at 64\,MHz. The Nordic nRF52832 is considered one of the most energy-efficient ARM Cortex-M4F on the market due to its low current consumption of only 46\,$\mu$W per MHz. Mr. Wolf has 64\,kB L1 memory closer to the cluster cores and 512\,kB L2 memory in the SoC domain. It can run up to 450\,MHz, with the most energy-efficient point being at 100\,MHz~\cite{pullini2018}, which has been used in this evaluation. Using the FPU, Network A executes in 38478 cycles, while in fixed point it needs 30210 cycles. The fixed-point implementation is 1.3$\times$ faster than the floating-point version and it is also more energy-efficient. Thus we will focus on the fixed-point implementation for the following analyses. The measured run-time in number of cycles for both Network A and Network B on the ARM Cortex-M4 and on Mr. Wolf is shown in Table~\ref{cycles} and the estimated energy consumption is listed in Table~\ref{energy}. The tables demonstrate three possible ways to use Mr. Wolf: 1) using only the SoC domain, i.e. the computation is done in the fabric controller (FC) using the IBEX core with the basic RV32IM instruction set and with the cluster powered off; or using the cluster to perform the computations on either either 2) a single RI5CY core, or 3) multiple RI5CY cores to do parallel computation. The RI5CY cores feature custom instruction set extensions to efficiently perform digital signal processing besides the basic RV32IM instruction set. These extensions offer 1.3$\times$ and 1.7$\times$ speed-up with respect to the ARM Cortex-M4 using CMSIS-DSP for Network A and B, respectively. However, the activation of the cluster domain costs more energy as shown in Table~\ref{energy}. The advantage of cluster computing in Mr. Wolf is much more evident when using multiple cores. The parallel execution using up to 8 cores provides respectively 4.9$\times$ and 8.3$\times$ speedup for Network A and B with respect to ARM Cortex-M4.


Besides the energy for the detection, it is essential to include the energy to acquire the data. Due to the low power analog front-end, the data acquisition of the ECG consumes only 171\,$\mu$W, while the GSR front-end consumes 30\,$\mu$W when active. To perform stress detection InfiniWolf, it acquires data for 3\,s (needing 600\,$\mu$J) and performs the feature extraction in 50\,$\mu$s (1\,$\mu$J assuming Mr. Wolf consuming 20\,mW in parallel execution~\cite{pullini2018}). The best overall energy cost for a single detection is 602.2\,$\mu$J thank to the extremely fast parallel execution of the classification.
\vspace{-0.15cm}
\subsection{Power Generation and Self-sustainability}\label{subSec:powerGen}
\begin{figure}[!t]
      \centering
      \includegraphics[trim={0cm 0cm 0cm 1cm},clip=true,width=2.5in]{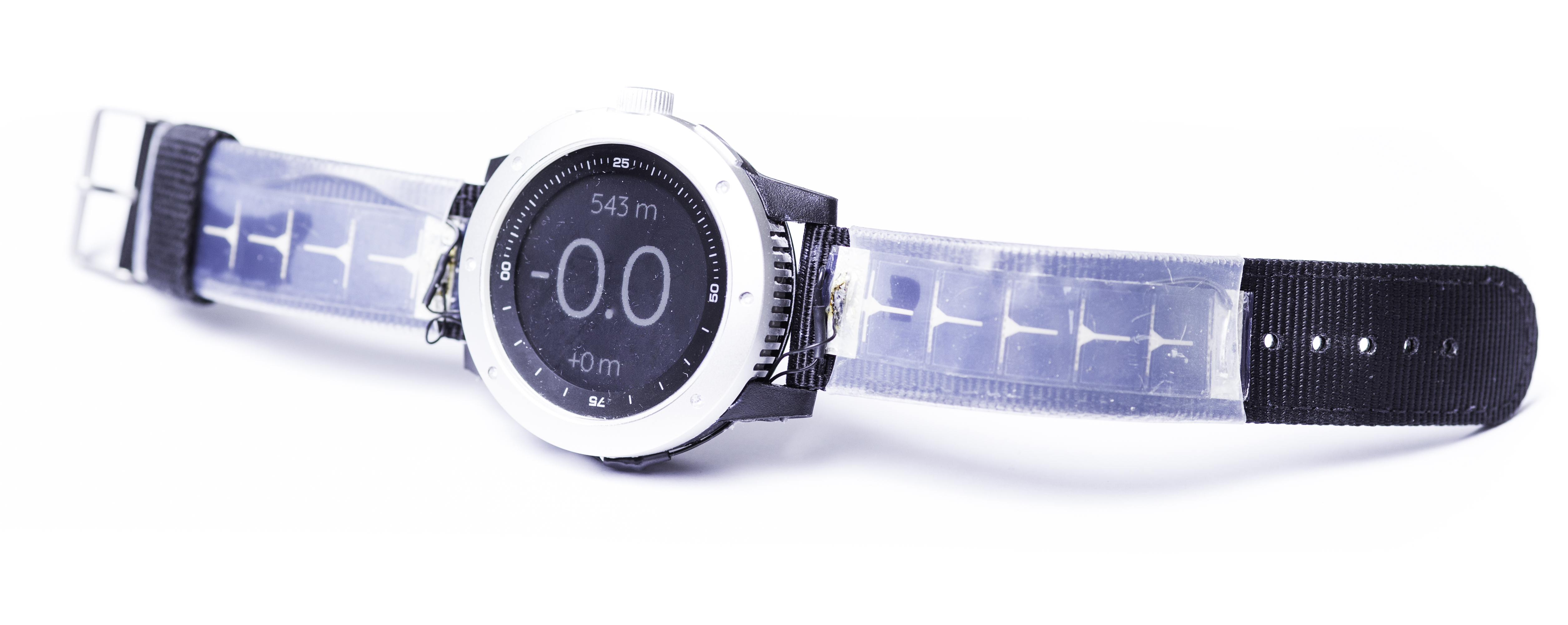}
      \vspace{-0.75cm}
      \caption{Prototype of InfiniWolf where the solar panels are visible. The TEG transducer is on the bottom to touch the human wrist where it is warmer. Desing based on \cite{Baumann19} }
      \vspace{-0.45cm}
      \label{fig:finalSystem}
\end{figure}

In this section, we provide information about the energy-harvesting capabilities of InfiniWolf, which hosts multiple energy harvesters (thermal and solar). The power generation of each source has been evaluated separately under laboratory conditions. Both harvesting sources have been attached to InfiniWolf into the Matrix Powerwatch case. InfiniWolf uses the built-in Matrix Industries TEG-module, while we adopted two small thin-film flexible photovoltaic panels SP3-12 by Flexsolarcells as shown in Fig.~\ref{fig:finalSystem}. We measured the power intake in our lab using the precise Keysight B2900A Series Precision Source/Measure Unit (SMU), which constitutes the heart of the measurement. 
InfiniWolf was sent to sleep mode, so the measurements take into account all the losses and the quiescent current of InfiniWolf.

Table~\ref{tbl:solarInAndOut} shows the measurements under different lighting conditions typical for indoor (700\,lx) and outdoor (30\,klx with sun) scenarios. 

\begin{table}[!b]
    \caption{Solar Power Generation Under Different Lighting Conditions}
    \vspace{-0.20cm}
      \centering
      \begin{tabular}{ |c| c| c| }
        \hline
        \textbf{ } & \textbf{Outdoor} & \textbf{Indoor}\\
        \hline
        Light Intensity: & \SI{30}{\kilo\lux} & \SI{700}{\lux}\\
        \hline
        Power Generation: & 24.711 mW & 0.9 mW\\
        \hline
      \end{tabular}
      \label{tbl:solarInAndOut}
      \vspace{-0.25cm}
\end{table}
 
In the same way, we measured the intake power into the battery. Table~\ref{tbl:tegWind} shows the measured data with different temperature gradients between the two TEG plates. Note that the TEG continuously generates energy in every condition. 
\begin{table}[!b]
    \caption{Human Wrist TEG Power Harvesting Measurements With and Without Active Cooling}
    \vspace{-0.20cm}
    \centering
    \begin{tabular}{|c|c|c|c|}
    \hline
    \textbf{Conditions:} & Room: \SI{22}{\celsius}     & Room: \SI{15}{\celsius} & Room: \SI{15}{\celsius} \\ \cline{2-4} 
                         & Skin: \SI{32}{\celsius}     & Skin: \SI{30}{\celsius} & Skin: \SI{30}{\celsius}  \\ \cline{2-4} 
                         & no wind                              & no wind                          & 42\,km/h wind      \\ \hline
    \textbf{Power:}      & \textbf{24.0\,$\mu$W} & \textbf{55.5\,$\mu$W} & \textbf{155.4\,$\mu$W}              \\ \hline
    \end{tabular}
    \label{tbl:tegWind}
    \vspace{-0.25cm}
\end{table}

Assuming InfiniWolf staying in challenging indoor condition for 6 hours, and the worst-case scenario for the TEG energy harvester, it will acquire 21.44\,J per day. As the intake energy is calculated including the losses and Infiniwolf quiescent current, the stress detection rate to achieve self-sustainability can be approximated as the total energy divided by the energy per detection calculated previously. Then Infiniwolf is able to perform up to 24 detection per minute, which is more than enough to many application scenarios.

\begin{table}[!b]
\caption{Runtime in Cycles, PULP v. ARM Cortex-M4F}
\vspace{-0.25cm}
\begin{center}
\begin{tabular}{|c|c|c|c|c|}
\hline
Runtime & ARM & \multicolumn{3}{c|}{PULP Mr. Wolf} \\ \cline{3-5} 
in \#cycles     &      Cortex-M4    & IBEX    & Single RI5CY & Multi RI5CY \\ \hline
Network A                                                                    & 30210                          & 40661 & 22772        & 6126        \\ \hline
Network B                                                                    & 902763                         & 955588   & 519354          & 108316         \\ \hline
\end{tabular}
\label{cycles}
\end{center}
\vspace{-0.25cm}
\end{table}
\vspace{-0.45cm}
\begin{table}[!b]
\caption{Energy Consumption per Classification [$\mu$J]}
\vspace{-0.45cm}
\begin{center}
\begin{tabular}{|c|c|c|c|c|}
\hline
Energy & Nordic w/ ARM & \multicolumn{3}{c|}{PULP Mr. Wolf} \\ \cline{3-5} 
in $\mu J$     & Cortex-M4    & IBEX    & Single RI5CY & Multi RI5CY \\ \hline
Net. A                                                                    & 5.1                          & 1.3 & 2.9        & 1.2        \\ \hline
Net. B                                                                    & 153.8                        & 31.5   & 65.6          & 21.6         \\ \hline
\end{tabular}
\label{energy}
\end{center}
\vspace{-0.25cm}
\end{table}

\section{Conclusion}
We presented InfiniWolf, an ultra-low-power, energy-efficient wearable device with multiple sensors able to run complex ML algorithms. The Mr. Wolf SoC allows having a long-term and even self-sustaining wearable device by combining dual source energy harvesting with extremely energy-efficient processing using the novel open-hardware parallel ultra-low-power processor based on RISC-V. Experimental results show the incredible energy-efficiency of the processor for a stress detection application using a multi-layer perception..



\bibliographystyle{IEEEtran}
\bibliography{IEEEabrv,bib}
\ifx

\fi

\end{document}